\documentstyle[prb,aps,epsf,eqsecnum]{revtex}

\begin{document}
\draft 

\begin{title}
{Mesoscopic Stoner Instability}
\end{title}
\author{I.L. Kurland$^{1,4}$, I.L. Aleiner$^{2,4}$ and 
B.L. Altshuler$^{1,3,4}$}
\address{$^{1}$Physics Department, Princeton University, Princeton, NJ 08544\\
$^{2}$Department of Physics and Astronomy, 
State University of New York, Stony Brook, NY 11794\\
$^{3}$NEC Research Institute, 4 Independence Way, Princeton, 
NJ 08540\\
$^{4}$Centre for Advanced Study, Drammensveien 78, N-0271 Oslo, Norway}
\date{\today}
\maketitle
\begin{abstract}
The paper is devoted to the magnetic properties of isolated mesoscopic
grains.  We demonstrate that under very general conditions
the electron - electron interactions in
such grains can be taken into account by
a simple interaction Hamiltonian. This Hamiltonian involves only three
coupling constants which correspond to charging, exchange
interaction, and superconducting correlations.  The most important
condition for such a description is that Thouless conductance of each
grain is large.  Under this condition sample-to-sample fluctuations of
the coupling constants can be neglected. However, the thermodynamic
properties can still remain sample-specific due to the one-electron
part of the Hamiltonian. If the grain is made from a material which 
is close to the
threshold of ferromagnetic instability the mesoscopic fluctuations
of the magnetization
are especially strong. Moreover, the situation becomes
multi-stable: free energy of each grain as a function of the
magnetization is characterized by a large number of local minima.  We
analyze the statistics of these minima and show that it possesses
simple scaling properties. Numerical simulations confirm this scaling.
\end{abstract}
\section{Introduction}
\label{sec:1}
The simplest model of ferromagnetism in metallic 
systems was proposed by Stoner\cite{4}. The magnetic ordering in this
model takes place when the increase in the orbital energy due to the 
promotion of 
electrons to higher energy states is smaller than the energy gain due to 
the exchange interaction.  As soon as this
 happens, the system becomes unstable 
with
respect to the transition to a state with a non-zero total spin
$S$ and, hence, broken ${\cal T}$-invariance.  In a bulk system only
macroscopically large $S$ matters. Therefore, the onset of the
instability is well determined - both the exchange and the orbital
energies are self-averaging quantities.

This might not be the case for a small (mesoscopic) metallic
grain since both orbital and exchange energies are sample specific.
Therefore, one should expect strong mesoscopic fluctuations of
the magnetization of the grain made from a material that is
close to the Stoner instability.  This paper is an attempt to
develop a theoretical description of these mesoscopic
fluctuations.

\begin{figure}[ht]
\vspace{0.15in}
\epsfxsize=6cm
\centerline{\epsfbox{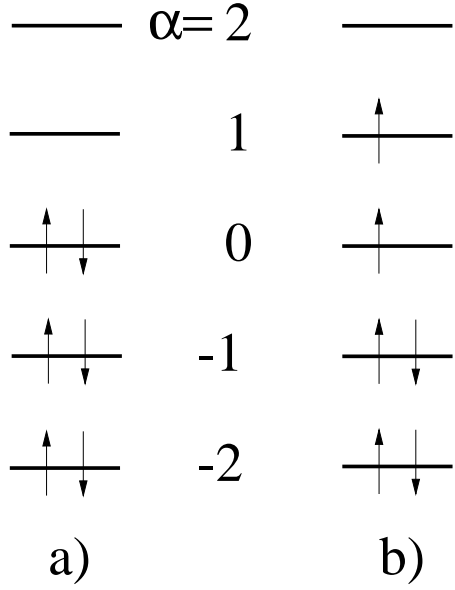}}
\vspace{0.15in}
\caption{Spin configuration for a) the lowest energy $S=0$ state. 
$\alpha=0,-1,-2,\dots$ correspond to $2$-orbitals. $\alpha=1,2,\dots$ 
correspond to $0$-orbitals; b) the lowest 
energy 
$S=1$ state. $\alpha=-1,-2,\dots$ correspond to $2$-orbitals, $\alpha=0,1$
correspond to $1$-orbitals, and  $\alpha=2,3,\dots$ correspond to $0$-orbitals.
}
\label{fig:1}
\end{figure}

Consider a mesoscopic grain with an even number of electrons at zero
temperature, T=0.  Restricting ourselves to the Hartree-Fock
approximation, we can speak about orbital states with energies
${\cal E}_{\alpha}$.  For a weak exchange interaction all of the
orbitals below the Fermi level $E_F$ are double-occupied (2-orbitals),
while those with ${\cal E}_{\alpha} < E_{F}$ are empty (0-orbitals).
This state is a singlet, $S=0$.  Let $\alpha$=0,1 label
correspondingly the highest 2-orbital and the lowest 0-orbital in
the singlet ground state, see Fig.~\ref{fig:1} 

For the stronger exchange a state with $S_{tot} \neq 0$ can become a
ground state. Indeed, let us compare the lowest energies $E_{0}(S)$ of
the singlet $(S=0)$ and triplet $(S=1)$ states (see Fig.~\ref{fig:1}).  
The lowest triplet
state has two single occupied orbitals: $\alpha=0,1$ (1-orbitals), while
$\alpha >1$ and $\alpha <0$ label 0-orbitals and 2-orbitals
respectively.  For the ferromagnetic sign of the spin exchange the two
electrons on the 1-orbitals have parallel spins.  
Assuming $SU(2)$ symmetry in the spin space of the system  we can write 
the exchange energy of two electrons, which occupy the orbital states 
$\alpha$ and $\beta$, in the usual Heisenberg form
\begin{equation}
\epsilon^{H}_{\alpha \beta}= -2 J_{\alpha \beta}\hat{\boldmath s}_1\hat{\boldmath s}_2, 
\label{Heisenberg}
\end{equation}
where $\hat{\mbox{\boldmath $s$}}_{1,2}$ are the spin operators. From this 
point on 
the superscript $H$ indicates the quantities belonging to the systems with 
the Heisenberg form of the exchange interaction, Eq.~(\ref{Heisenberg}).  
We can write the following expression for the energy difference between the
lowest energy states with $S=0$ and $S=1$,
\begin{equation}
E_{0}^H (1) - E_{0}^H (0) = {\cal E}_1-{\cal E}_0 - \frac{1}{2}J_{01} - 
\frac{3}{2}J_{00}. 
\label{1}
\end{equation}
[In derivation of Eq.~(\ref{1}) we used the fact that 
$2\hat{\mbox{\boldmath $s$}}_1\hat{\mbox{\boldmath $s$}}_2=
S(S+1)-3/2$, when we add spins of two $S=1/2$ particles].  Therefore,
the triplet state becomes energetically more favorable than the singlet one,
 $E_{0}(1) < E_{0}(0)$, provided that 
$J_{01}+3J_{00}~>~2({\cal E}_1-{\cal E}_0)$.

This is a sufficient rather than a necessary condition for $S$ to be nonzero
in the ground state.  Even for $E_{0}(1)>E_{0}(0)$, the spin $S$ of
the ground state is not necessarily zero.  For the lowest energy
among the states with a total spin $S$ we have  
\begin{equation}
E^{H}_{0}(S) - E^{H}_{0}(0) = \sum^{S}_{\alpha = 1} 
\left(
{\cal E}_{\alpha} -
{\cal E}_{1-\alpha} 
- \frac{3}{2}J_{1-\alpha, 1-\alpha}
\right) 
- \sum_{1-S \leq \alpha<
  \beta \leq S} \frac{1}{2}J_{\alpha \beta}.
\label{energyH}
\end{equation}
Here $\alpha < 1-S$ labels 2-orbitals,
$1-S \leq \alpha \leq S$ labels 1-orbitals and 0-orbitals have labels $\alpha > S$.

One can also consider situation with strong spin anisotropy. 
Let it be the easy axes (Ising) anisotropy, so that the state of an 
electron
is characterized by the $z$-component $s_z$ of its spin ($s_z=\pm 1/2$), and 
the exchange energy equals to
\begin{equation}
\epsilon^{I}_{\alpha \beta}= -2 J_{\alpha \beta}s_{z(1)}s_{z(2)}. 
\label{Ising}
\end{equation}
From this point on 
the superscript $I$ indicates the quantities belonging to the systems with 
the Ising form of the exchange interaction (Eq.~(\ref{Ising})).
The energy difference between states with zero and finite total spin in this 
Ising case is
\begin{equation}
E^{I}_{0}(S) - E^{I}_{0}(0) = \sum^{S}_{\alpha = 1} 
\left(
{\cal E}_{\alpha} -
{\cal E}_{1-\alpha} 
-\frac{1}{2} J_{1-\alpha, 1-\alpha}
\right) 
- \sum_{1-S \leq \alpha,
  \beta \leq S}\frac{1}{2} J_{\alpha \beta}.
\label{energyI}
\end{equation}

In a general case the orbital energies ${\cal E}_{\alpha}$ and
the exchange energies $J_{\alpha \beta}$ are random.  In the limit $S
\rightarrow \infty$ summation in Eqs.~(\ref{energyH}), (\ref{energyI}) 
leads to the self-averaging.  
Calling $\delta_{1} = \langle {\cal E}_{\alpha +1} - {\cal E}_{\alpha}
\rangle$ mean level spacing and $J = \langle J_{\alpha \beta}\rangle$ mean
exchange coupling constant
(here $\langle ... \rangle$ stands for the ensemble averaging), we obtain from
Eqs.~(\ref{energyH}) and (\ref{energyI})
\begin{mathletters}
\begin{eqnarray}
E_0^{H}(S) - E_0^{H}(0) = S^{2} \delta_{1} -\frac{1}{2} \left[3S + \frac {2S(2S-1)}{2}\right] J =
S^{2}\delta_{1} -JS(S+1);\label{instability.h}\\
E_0^{I}(S) - E_0^{I}(0) = S^{2} \delta_{1} -\frac{1}{2} \left[S + \frac {2S(2S-1)}{2}\right] J =
S^{2}\delta_{1} -JS^{2},
\label{instability}
\end{eqnarray} 
\end{mathletters}
where $S$ is an integer.

Note that for a grain with an odd number of electrons the 
Eqs.~(\ref{instability.h}) and (\ref{instability}) are slightly modified. 
The lowest possible spin is now $S=1/2$ and the energy of a state with a 
spin $S$ is given by
\begin{mathletters}
\begin{eqnarray}
E_0^{H}(S) - E_0^{H}\left(\frac{1}{2}\right) =
S^{2}\delta_{1} -JS(S+1)-\frac{1}{4}\delta_{1}+\frac{3}{4}J;
\label{instabilityodd.h}\\
E_0^{I}(S) - E_0^{I}\left(\frac{1}{2}\right) =
S^{2}\delta_{1} -JS^{2}-\frac{1}{4}\delta_{1}+\frac{1}{4}J,
\label{instabilityodd}
\end{eqnarray}
\end{mathletters} 
where $S$ is a half integer.
One can see that for both Heisenberg and Ising exchange interactions, 
the system is unstable at
\begin{equation}
\delta_{1} < J.
\label{Stoner}
\end{equation}
Indeed, under this condition $E(S)$ in Eq.~(\ref{instability}) tends to 
$-\infty$ 
as $S\rightarrow \infty$. 
Equation (\ref{Stoner}) is nothing but the familiar Stoner criterium of this 
instability. As soon as the parameters of the system surpass the instability 
threshold,
the ground state acquires magnetization proportional to the volume of the 
system.

The situation in finite systems is somewhat different.
We start our discussion with a simple but instructive example of a grain
with an even number of electrons without any disorder, i. e.  
${\cal E}_{\alpha +1} - {\cal E}_{\alpha}=\delta_{1}$ and $J_{\alpha \beta}=J$.
It turns out that for the Heisenberg case the spin of the ground state $S_g$ 
is finite already at
\begin{equation} 
\delta_1<2J, 
\end{equation}
as can be directly seen from Eq.~(\ref{1}). 
Therefore, $S$ can be greater than zero even for
the exchange which is below the critical value given by the Stoner
criterion (\ref{Stoner}) ($\delta_1/2=J_0<J<J_c=\delta_1$). In this 
parameter domain
$S_g=J/(2\delta_1-2J)$. Because $\delta_{1}$ and $J$
are inversely proportional to the volume of the system, the spin of the ground 
state $S_g$
does not scale with the volume. Equation~(\ref{instabilityodd.h}) implies that 
for an odd number of electrons the system spin $S_g>1/2$ (though finite) in 
the interval
$2\delta_1/3=J_0~<~J~<~J_c=\delta_1$.
On the contrary, in the case of Ising interaction
$S_g=0$ ($1/2$) for an even (odd) number of electrons as long 
as $\delta_{1}~>~J$.

In more realistic models, which take the randomness in both the level
spacing and the exchange interaction
into account, the magnetization at $0~<~J~<~\delta_1$ is 
{\em essentially random}. 
Situation far from Stoner instability was discussed in
Refs.~\onlinecite{13,14}. 
In this paper we propose a theory of the magnetization near this 
critical point, $\delta_1-J\ll\delta_1$. 

 For finite $S$ the functions $E_{0}(S)$ from Eqs.~(\ref{instability.h})
and (\ref{instability}) are random. Their
statistics are determined by the probability distributions of
$E_{\alpha}$ and $J_{\alpha \beta}$.  Below we derive the correlation
function $\langle E_{0} (S_{1}) E_{0}(S_{2})\rangle$ for large but
finite $S_{1,2}$ in a realistic model of a weakly disordered or
chaotic metallic grain.  This correlation function is sufficient to
describe the statistical mechanics of the grain. At high temperatures the
randomness can be treated perturbatively in powers of $1/T$ (high 
temperature expansion).  
The low-temperature behavior of those systems is determined by the 
position and
the structure of the deepest
minimum of the $E(S)$-function.  Therefore, one has to develop a 
description of the spin $S_{g}$ of the ground state: $E_{0}(S_{g})=
  \min \{E_{0}(S)\}$.  Since $S_{g}$ is random, one might be
interested in the statistics of $S_{g}$ and of $E(S)$ for $S$ close to $S_{g}$.
The problem of evaluation of statistics of $S_g$ is not trivial. In this 
paper we do not solve it completely. We restrict
ourselves to scaling analysis and numerical simulation. Our attempt to
construct an analytical description based on the Replicas Symmetry
Breaking paradigm\cite{3} will be published elsewhere\cite{elsewhere}.

The remainder of this paper is organized in the following way.  In
 section~\ref{sec:2} we discuss a model of interacting electrons in a
weakly disordered metallic grain and the statistics of $E_{\alpha}$
and $J_{\alpha\beta}$, that follow from this model.
Section~\ref{sec:3} is devoted to the derivation of the correlation
function $\langle E_{0}(S_{1}) E_{0} (S_{2})\rangle$. The scaling
analysis of the structure of the minima of $E_{0}(S)$ is performed in
Sec.~\ref{sec:4}.  In Sec.~\ref{sec:5} the results are compared with numerical
 simulations. Our findings are summarized in  the Conclusion.

\section{ Electron-Electron Interactions in Isolated Metallic Grains}
\label{sec:2}

Let us discuss what the energy scales that determine
properties of a finite system of electrons are.  The single electron
spectrum $\{{\cal E}_{\alpha}\}$ is characterized by the mean level
spacing
\begin{equation}
\delta_{1} = \langle {\cal E}_{\alpha + 1} - {\cal E}_{\alpha} \rangle,
\end{equation}
where $\langle ... \rangle$ stands for the ensemble 
averaging.  Another relevant energy scale of the problem is the Thouless
energy $E_T \approx \hbar/t_{erg}$, where $t_{erg}$ is the time it takes for
a classical counterpart of an
electron to cover the energy shell in the single-particle phase
space.  For diffusive and ballistic systems $t_{erg}$ equals to
$L^{2}/D$ and $L/v_F$ respectively.  Here $L$ is the size of the
system, $v_{F}$ denotes the electronic Fermi velocity, and $D$ is the
diffusion coefficient.

The  important characteristic of the system is the ratio of
these two energy scales
\begin{equation}
g = E_{T}/\delta_{1},
\end{equation}    
which is known as the dimensionless conductance.  Here
we discuss only metallic grains where all single-electron states
are extended, and, thus, the dimensionless conductance is 
large.
\begin{equation}
g\gg 1.
\label{eq:2.3}
\end{equation}

It is well known \cite{12,16,15} that in this regime the statistics
of the spectrum $\{{\cal E}_{\alpha}\}$ on the scales smaller than the
Thouless energy $E_{T}$ are well described by Random Matrix
Theory (RMT) \cite{9,10,11}.  RMT gives a quantitative 
description of the phenomenon of the level repulsion.  For an ensemble of
$N \times N, N \rightarrow \infty$ matrices with random and
independent matrix elements the probability density of a
realization of the spectrum $\{{\cal E}_{\mu}\}$ is given by\cite{9,11}:
\begin{equation}
P(\{{\cal E}_{\mu}\}) \propto \exp \left[\frac{\beta}{2} \sum_{\mu \neq \nu}\ln
\left(\frac  {\vert {\cal E}_{\mu} - {\cal E}_{\nu} \vert}{\delta_{1}}\right)\right].
\label{eq:2.1}
\end{equation} 
The parameter $\beta$ in Eq.~(\ref{eq:2.1}) can be equal to $1$, $2$ or $4$ for orthogonal,
unitary and simplectic ensembles respectively.  The orthogonal
(unitary) RM ensemble corresponds to weakly disordered grains with
preserved (violated) ${\cal T}$-invariance and negligible interaction
between orbital and spin degrees of freedom.  In what follows, we
restrict ourselves by these two ensembles, i.e., do not consider
grains with spin-orbit interaction. The latter would correspond
to the simplectic ensemble, $\beta=4$. 

Now let us take the effects of electron-electron
interaction in the grain into consideration.  It turns out that a large class 
of such
effects in a {\it given} disordered closed metallic grain can, under
very general conditions, be described by a remarkably simple
Hamiltonian with only three system dependent coupling
constants. Let the dimensionless conductance 
$g$ of the grain tend to infinity. We
start with the simplest case when the electrons interact via a
short-range potential 
\begin{equation}
\hat{H}_{int}(\vec{r})=\lambda\delta_1 V\delta(\vec{r}). \label{v.int}
\end{equation} 
Here $V\propto L^d$ is the volume of the grain and
$\lambda$ is a dimensionless coupling constant. The matrix 
element of
this interaction in the basis of eigenstates $\varphi_{\alpha}(\vec{r})$
of the noninteracting Hamiltonian is given by
\begin{equation}
M^{\alpha \gamma}_{\mu \nu}=\lambda\delta_1 V\int d\vec{r}\varphi_{\alpha}^{\ast}
(\vec{r})\varphi_{\gamma}^{\ast}(\vec{r})\varphi_{\mu}(\vec{r})\varphi_{\nu}(\vec{r}).
\label{mtx1}
\end{equation} 

We first consider the situation without magnetic field (${\cal
T}$-reversal invariance is preserved). Since no spatial symmetries are
assumed, {\em the one particle orbitals are not degenerate} and the
eigenfunctions of one-particle Hamiltonian $\varphi_{\mu}(\vec{r})$
{\em can be chosen to be real}.  The off-diagonal elements in
Eq.~(\ref{mtx1}) are small as $\delta_1/E_T=1/g$ because the integrand
quickly oscillates. Thus, they can be neglected. On the contrary, diagonal 
matrix
elements ($\alpha$, $\gamma$, $\mu$, $\nu$ are equal pairwise) are much
larger since the integrand in Eq.~(\ref{mtx1}) is positive
definite. Substituting the integrand by its mean value, e. g.,
\begin{equation}
\langle \varphi_{\alpha}^{2}(\vec{r})\varphi_{\gamma}^{2}(\vec{r})\rangle=
V^{-2},
\end{equation} 
we find
\begin{equation}
M^{\alpha \gamma}_{\alpha \gamma}=M^{\alpha \alpha}_{\gamma \gamma}=
M^{\alpha \gamma}_{\gamma \alpha}=\lambda\delta_1.\label{mtx2}
\label{melements}
\end{equation}
Corrections to Eq.~(\ref{mtx2}) are negligible for exactly the same
reason the off-diagonal matrix elements are. As a result, in the
limit $E_T\rightarrow \infty$, Eq.~(\ref{mtx2}) becomes exact!

For the following discussion it is convenient to introduce operators
of the number of electrons $\hat{n}_{\alpha}$ on the orbital $\alpha$
and the spin $\vec{S}_\alpha$ on this orbital
\begin{equation}
\hat{n}_{\alpha}=\sum_{\sigma}a_{\alpha\sigma}^{\dag}a_{\alpha\sigma}; \quad 
\vec{S}_\alpha=\frac{1}{2}\sum_{\sigma\sigma_1}a_{\alpha\sigma}^{\dag}
a_{\alpha\sigma_1}\vec{\sigma}_{\sigma\sigma_1}.\label{ns}
\end{equation} 
Here $a_{\alpha\sigma}^{\dag}$ ($a_{\alpha\sigma}$) creates
(annihilates) an electron with a spin $\sigma$ on the orbital $\alpha$
and $\sigma_{\sigma\sigma_1}^i$ are the Pauli matrices. Neglecting
the off-diagonal terms in the interaction Hamiltonian reduces it to
the form
\begin{eqnarray}
&\hat{H}_{int}=\hat{H}^{(1)}+\hat{H}^{(2)}+\hat{H}^{(3)}=&\label{hint}\\
&\sum_{\alpha\beta\sigma\sigma_1}\left[ 
M^{\alpha \gamma}_{\alpha \gamma}a_{\alpha\sigma}^{\dag}a_{\alpha\sigma}
a_{\gamma\sigma_1}^{\dag}a_{\gamma\sigma_1}+
M^{\alpha \gamma}_{\gamma \alpha}a_{\alpha\sigma}^{\dag}a_{\alpha\sigma_1}
a_{\gamma\sigma_1}^{\dag}a_{\gamma\sigma}
+M^{\alpha \alpha}_{\gamma \gamma}a_{\alpha\sigma}^{\dag}a_{\alpha\sigma_1}^
{\dag}a_{\gamma\sigma}a_{\gamma\sigma_1} \right].&\nonumber
\end{eqnarray}
To begin with, let us consider the first two terms in Eq.~(\ref{hint}). Using
Eq.~(\ref{ns}) and the identity
\[
\vec{\sigma}_{\sigma\sigma '}\vec{\sigma}_{\sigma_1\sigma_1 '}=
2\delta_{\sigma\sigma_1 '}\delta_{\sigma '\sigma_1}-\delta_{\sigma\sigma '}
\delta_{\sigma_1\sigma_1 '}
\]
one can present $\hat{H}^{(1)}+\hat{H}^{(2)}$ in a usual Hartree-Fock form:
\begin{equation}
\hat{H}^{(1)}+\hat{H}^{(2)}=\sum_{\alpha \gamma}\left[ \left(
M^{\alpha \gamma}_{\alpha \gamma}-\frac{1}{2}M^{\alpha \gamma}_{\gamma \alpha}
 \right)\hat{n}_{\alpha}\hat{n}_{\gamma}-2M^{\alpha \gamma}_{\gamma \alpha}
\hat{\vec{S}}_{\alpha}\hat{\vec{S}}_{\gamma}\right].\label{hf}
\end{equation}
Now we can use the remarkable independence of the matrix elements of their 
indices, Eq.~(\ref{mtx2}), and present Eq.~(\ref{hf}) as
\begin{equation}
\hat{H}^{(1)}+\hat{H}^{(2)}=\lambda\delta_1 \left[\frac{1}{2}\hat{n}^2-
2\hat{\vec{S}}^2 \right].\label{16}
\end{equation}
Here $\hat{\vec{S}}$ and $\hat{n}$ are the operators of the total spin and the 
total number of electrons correspondingly,
\begin{equation}
 \hat{\vec{S}}=\sum_{\alpha}\hat{\vec{S}}_{\alpha}; \quad \hat{n}=
\sum_{\alpha}\hat{n}_{\alpha}.\label{n}
\end{equation}
The third term in Eq.~(\ref{hint}) can be treated in a similar fashion. One
can write it as 
\begin{equation}
\hat{H}^{(3)}=\lambda\delta_1\hat{T}^{\dag}\hat{T}; \quad \hat{T}=
\sum_{\alpha}\hat{a}_{\alpha\uparrow}\hat{a}_{\alpha\downarrow}. \label{18}
\label{T}
\end{equation}

It follows from Eqs.~(\ref{16}) and (\ref{18}) that the Hamiltonian
$\hat{H}_{int}$ can be presented in terms of three operators
$\hat{n}$, $\hat{S}$, and $\hat{T}$ from Eqs.~(\ref{n}), (\ref{18}),
rather that in terms of all the operators $\hat{n}_{\alpha}$,
$\hat{S}_{\alpha}$ and
$\hat{a}_{\alpha\uparrow}\hat{a}_{\alpha\downarrow}$ with different
$\alpha$. In fact, this feature is not specific to the particular
short range interaction Hamiltonian (\ref{v.int}), and is determined by 
the
chaoticity of the eigenfunctions $\varphi_{\alpha}(\vec{r})$.  It
turns out that $\varphi_{\alpha}(\vec{r})$ are Gaussian random
variables which are not correlated with each other. This is 
correct for eigenvectors of $N\times N$, $N\to\infty$ random 
matrices\cite{11}:
\begin{mathletters}
\label{corrfunctions}
\begin{equation} 
  \langle \varphi^{\ast}_{\mu}(m) \varphi_{\nu}(n)\rangle =
  \frac{1}{N} \delta_{\mu \nu}\delta_{mn}; \quad
 \langle \varphi_{\mu}(m)
  \varphi_{\nu}(n)\rangle = \frac{2-\beta}{N} \delta_{\mu \nu}\delta_{mn},
\label{19a}
\end{equation}
and $\langle \varphi_{\mu}(m)\rangle = 0$. It is also
correct for the Berry ansatz\cite{Berry} for the wavefunctions in a
chaotic grain
\begin{equation} 
  \langle \varphi^{\ast}_{\mu}(\vec{r_1})
  \varphi_{\nu}(\vec{r_2})\rangle = \frac{1}{V} \delta_{\mu \nu}
  h(\vec{r_1}-\vec{r_2}); \quad \langle \varphi_{\mu}(\vec{r_1})
  \varphi_{\nu}(\vec{r_2})\rangle = \frac{2-\beta}{V} \delta_{\mu
  \nu}h(\vec{r_1}-\vec{r_2}).\label{19b}
\end{equation}
\end{mathletters}
Here $h(\vec{r})$ is the familiar Friedel function, 
\begin{equation}
h(\vec{r})=\Gamma\left( d/2\right)\frac{J_{\frac{d}{2}-1}(x)}
{x^{\frac{d}{2}-1}};\quad x=\frac{2\pi\left|\vec{r}\right|}{\lambda_F},
\end{equation}
where $d$ is the spatial dimension of the grain, $J$ is the Bessel function and
$\lambda_F$ is the Fermi wavelength. 
The characteristic
scale of the decay of this function is
of the order of the electron wavelength, $\lambda_F$, and for the
purposes of convolution with any smooth function can be substituted by
$\lambda_F^3\delta(\vec{r})$. Equation (\ref{19b}) becomes exact in the limit
$g\rightarrow \infty$.

It is easy to see that the correlators of the eigenvectors (eigenfunctions)
(\ref{corrfunctions}) are invariant with respect to an arbitrary orthogonal
transformation performed over them:
\begin{mathletters}
\label{transformation}
\begin{eqnarray}
\varphi_\mu(m) \to \sum_{\mu^\prime, m^\prime}
O_{mm^\prime}^{\mu\mu^\prime}\varphi_{\mu^\prime}(m^\prime);
\quad 
\varphi_\mu(\vec{r}) \to \sum_{\mu^\prime}\int d \vec{r}_1
O^{\mu\mu^\prime}\left(\vec{r},\vec{r}_1\right)
\varphi_\mu(\vec{r}_1); \\ 
\sum_{\mu^\prime, m^\prime}
O_{mm^\prime}^{\mu\mu^\prime}O_{m^\prime n}^{\mu^\prime\nu}
=\delta_{mn}\delta_{\mu\nu}
;
\quad \int d \vec{r} \sum_{\mu^\prime}
O^{\mu\mu^\prime}\left(\vec{r}_1,\vec{r}\right)
O^{\mu^\prime\nu}\left(\vec{r},\vec{r}_2\right)
=\delta_{\mu\nu}\delta\left(\vec{r}_1 - \vec{r}_2\right)
.
\end{eqnarray}
Therefore, the ensemble averaged part of the interaction Hamiltonian
must also be invariant with respect to such transformations.  [As to
mesoscopic fluctuations of $\hat{H}_{int}$, they disappear in the
universal (RM) limit, $g\to \infty$ $(N\to\infty)$, and can be
neglected under the condition (\ref{eq:2.3}).] There are only three
operators, quadratic in the fermionic fields, which possess this invariance:
\end{mathletters}
\begin{eqnarray}
& {\hat{n}} = \sum_{\alpha,\sigma} \hat{a}^{\dagger}_{\alpha,\sigma}
\hat{a}_{\alpha \sigma};&\nonumber\\
& \hat{\vec{S}} = \frac{1}{2} \sum_{\alpha,\sigma_{1},\sigma_{2}}
\hat{a}^{\dagger}_{\alpha \sigma_{1}} \vec{\sigma}_{\sigma_{1},\sigma_{2}}
    \hat{a}_{\alpha,\sigma_{2}}; & \nonumber\\
&{\hat{T}} = \sum_{\alpha}\hat{a}_{\alpha\uparrow}\hat{a}_{\alpha\downarrow},
&\label{invariants}
\end{eqnarray}
and the quartic operators may be constructed only as  second powers
or products of these operators. Moreover, the Hamiltonian $\hat{H}_{int}$ 
should also be invariant with
respect to $SU(2)$ rotations in the spin space. Therefore, the spin may
enter into the interaction Hamiltonian only through the combination 
$\hat{\vec{S}}^2$. Finally,
$\hat{H}_{int}$ must be invariant with respect to $U(1)$ gauge
transformations. It means that $\hat{H}_{int}$ can include the operator
$\hat{T}$ only as a product $\hat{T}^\dagger\hat{T}$.
We conclude that in the general case the limit
of the infinite conductance $g$ corresponds to
\begin{equation} 
\hat{H}_{int}=E_c
\hat{n}^2-J(\vec{S})^2+\lambda_{BCS}\hat{T}^{\dagger}\hat{T}.
\label{Eq1}
\label{20}
\label{eq:2.17}
\end{equation}
(Linear in $\hat{n}$ term is allowed,  but it can be included into
the one-particle part of the Hamiltonian.)
We introduced three coupling constants $E_c$, $J$ and $\lambda_{BCS}$, that 
correspond to the three operators permitted by the symmetries. In the simple 
model with the short range interaction and preserved ${\cal T}$ -
invariance,
$\beta =1$, the above coupling constants have the following form:
\begin{equation} 
E_c=\frac{1}{2}\lambda\delta_1; \quad J=2\lambda\delta_1;\quad 
\lambda_{BCS}=\lambda\delta_1.
\end{equation}
If the ${\cal T}$ - invariance is broken, $\beta=2$, transformations
(\ref{transformation}) become unitary. The operator $\hat{T}$ from
Eq.~(\ref{invariants}) is incompatible with such symmetry and, thus,
$\lambda_{BCS}=0$, for $\beta=2$.

We would like to emphasize once again that Eq.~(\ref{20}) is the most 
general form
of the Hamiltonian in the limit $g\rightarrow \infty$. For instance, one 
can 
check that it correctly takes into account interactions of the two electrons
on the same orbital. Indeed, it follows from Eq.~(\ref{mtx1}) and
Gaussian distribution of $\varphi_\alpha(\vec{r})$ [with correlation
functions determined by Eqs.~(\ref{corrfunctions})], that the
``double diagonal'' matrix element $M_{\alpha\alpha}^{\alpha\alpha}$
is different from diagonal ones, given by Eq.~(\ref{melements}):
\begin{equation}
M_{\alpha\alpha}^{\alpha\alpha}=(4-\beta) 
M_{\alpha\gamma}^{\alpha\gamma} =(4-\beta)\lambda\delta_1.
\label{doublediagonal}
\end{equation}

On the first sight it appears that the Hamiltonian (\ref{Eq1}) has to
be supplemented by an additional term proportional to
$M_{\alpha\alpha}^{\alpha\alpha}$ \cite{14}:
\begin{equation}
\hat{H}_? \propto \lambda\delta \sum_\alpha 
\hat{a}^\dagger_{\alpha,\uparrow}\hat{a}_{\alpha,\uparrow}
\hat{a}^\dagger_{\alpha,\downarrow}\hat{a}_{\alpha,\downarrow}.
\label{wrong}
\end{equation}
However, one notices that for any two distinct labels $\alpha\neq \gamma$ 
there are three different off-diagonal elements
$M_{\alpha\gamma}^{\alpha\gamma}$, $M_{\alpha\gamma}^{\gamma\alpha}$,
and $M_{\alpha\alpha}^{\gamma\gamma}$, whereas
double-diagonal term $M_{\alpha\alpha}^{\alpha\alpha}$ should be taken
into account only once. As a result, the relationship
(\ref{doublediagonal}) is exactly what is needed for Eq.~(\ref{hint})
to be valid, provided that the sum in the right-hand side involves
terms with $\alpha=\gamma$ as well as those with $\alpha\neq \gamma$.
Therefore, the interaction between electrons on the same orbital {\em
does not violate the invariance under the rotation of the basis},
which becomes exact in $g \to \infty$ limit. On the other hand, {\it an
additional term of the form Eq.~(\ref{wrong}) explicitly violates this
symmetry}. For this reason we think that taking such a term into account,
as it was done in Ref.~\cite{14} is incorrect.

Let us briefly discuss the physical meaning of the three terms in
the Hamiltonian Eq.~(\ref{Eq1}).  The last one exists only in the
orthogonal case $(\beta = 1)$ and leads to the superconducting
instability provided that $\lambda_{BCS} < 0$, i.e., there is
an attraction in the Cooper channel. Superconducting correlations are
suppressed by the magnetic field, and thus do not exist at $\beta=2$.  
Here we are not interested in the
effects of superconductivity and assume that the grain is a
normal metal at $T=0$. It means that  $\lambda_{BCS} > 0$.  It is
well known that the very same renormalization which leads to
superconductivity at $\lambda_{BCS} > 0$ renormalizes the negative effective
coupling constant in the Cooper channel to zero (see, e.g., 
Ref.~\onlinecite{17}). 
This fact enables us to simply neglect this interaction.

The first two terms in Eq.~(\ref{Eq1}) represent the dependence of the
energy of the grain on the total number of the electrons inside
and on the total spin respectively.  They commute with each
other and with the single-particle part of the Hamiltonian $\sum
{\cal E}_{\alpha} \hat{a}^{\dagger}_{\alpha,\sigma} 
\hat{a}_{\alpha,\sigma}$ provided the grain is isolated. 
Therefore, all states of the
  grain can be classified by $n$ and $S$.

As long as we are interested in spin structure of the system with a
{\it fixed number of electrons}, we can disregard the first term in
Eq.~(\ref{Eq1}).  As a result we arrive to a simple Hamiltonian
\begin{equation}
H = \sum_{\alpha,\sigma} {\cal E}_{\alpha}
\hat{a}^{\dagger}_{\alpha,\sigma}\hat{a}_{\alpha,\sigma} - J({\hat{S}})^{2}.
\end{equation}

Note that the only random component of the problem is
the single-particle spectrum $\{{\cal E}_{\alpha}\}$, while the exchange
$J$ does not fluctuate!

It should be noted though, that RM theory is just an
approximation.  It becomes exact only in the limit $g\to\infty$. For finite $g$
there are corrections to Eq.~(\ref{Eq1}), which sometimes are of
importance.  However, these corrections at $g\gg 1$ do not bring
essentially new physics to the problem of the small grain
magnetization.  For this reason we restrict ourselves to the
Hamiltonian Eq.~(\ref{Eq1}).

Choosing the direction of the total spin of the system to
coincide with the $z$-axis, we can express the energy of 
the system through the occupation numbers
$n_{\alpha,\sigma}$:
\begin{equation}
E^H=\sum_{\alpha,\sigma} {\cal E}_{\alpha} n_{\alpha,\sigma} - J S(S+1)\label{h1},
\label{eq:2.20}
\end{equation}
where
\begin{equation}
S = \frac{1}{2}\left|\sum_{\alpha} (n_{\alpha,\uparrow} - n_{\alpha,
  \downarrow })\right| = |S_{z}|,
\label{eq:2.21}
\end{equation}
and $\sum n_{\alpha,\sigma} = N$.

Throughout this section we assumed that the metallic grain possesses 
rotational symmetry in the spin space. In addition, 
it is worthwhile to discuss 
a case in which this symmetry 
is broken and the interaction as only along one easy axes. To make 
connection with the energy $E_0^I(S)$ that has been defined in 
Eq.~(\ref{instability}) we adopt the following Hamiltonian:
\begin{equation}
H = \sum_{\alpha,\sigma} {\cal E}_{\alpha}
\hat{a}^{\dagger}_{\alpha,\sigma}\hat{a}_{\alpha,\sigma} - J({\hat{S_z}})^{2}.
\end{equation}  
Using the notation of Eq.~(\ref{eq:2.21}) we express the energy for the 
Ising case as
\begin{equation}
E^I=\sum_{\alpha,\sigma} {\cal E}_{\alpha} n_{\alpha,\sigma} - J S^2.
\label{eq:2.22}
\end{equation}

The energy $E_{0}(S)$ which was introduced in the previous section can
be obtained by minimizing $E$ at fixed $S$ with respect to the
occupation numbers $n_{\alpha,\sigma}$.  In spite of the simple form
of Eqs.~(\ref{eq:2.20}), (\ref{eq:2.22}) and (\ref{eq:2.21}) the problem 
remains
non-trivial, since the spectrum $\{{\cal E}_{\alpha}\}$ is random.  In
the following sections
we consider the effect of this randomness on the properties of the
ground state.

\section{The Effective Hamiltonian}
\label{sec:3}
 
We are interested in the spontaneous magnetization of the system at low
temperatures. Therefore, in the following discussion we only compare
the energies of the lowest lying states with a given spin. Hence, the
state labeled by the total spin $S$ is the state for which the first
$N/2+S$ single electron states are occupied by spin up and the first $N/2-S$
by spin down electrons. We ignore the excited states with
the same total spin in the following discussion. (It can be shown
that taking these states into account
introduces a nonessential addition to the free energy). This enables us
[based on the Hamiltonian (\ref{h1})] to write the following energy
functional:
\begin{equation}
E(S,\xi)=\int^{{\cal E}_>}_0 \epsilon \rho(\epsilon)d\epsilon
-
\int^0_{{\cal E}_<}\epsilon \rho(\epsilon)d\epsilon
-JS(S+\xi-1)),
\label{eq:3.1}
\label{IN3}
\end{equation}
where $\rho(\epsilon)=\sum_\alpha \delta\left(\epsilon-{\cal E}_\alpha\right)$
 is the one electron density of states and $J$ is the
strength of the exchange interaction, and parameter $\xi=1,2$
characterizes two limiting cases of spin anisotropy: $\xi =2$
corresponds to the isotropic spin orientation (Heisenberg model) with
the degeneracy of the state $2S+1$, while $\xi =1$ describes the easy
axes (Ising) model where states are only double degenerate. 
We have $2$-orbitals at ${\cal
E}_{\alpha}<{\cal E}_<$, $1$-orbitals at ${\cal E}_< <{\cal E}_\alpha
<{\cal E}_>$ and empty $0$-orbitals at ${\cal E}_{\alpha}>{\cal
E}_>$. From the conservation of the total number of particles 
we obtain the following equations for ${\cal E}_>$ and ${\cal E}_<$
\begin{equation}
S=\int^{{\cal E}_>}_0 \rho(\epsilon)d\epsilon=
\int^0_{{\cal E}_<}\rho(\epsilon)d\epsilon.
\label{IN4}
\label{eq:3.2}
\end{equation}
The minimum of the energy functional (\ref{eq:3.1}) is the same as the ground 
state
of the original system. The density of states can be represented as
\begin{equation}
\rho(\epsilon)=1/\delta_1+\delta\rho(\epsilon). 
\end{equation}
We are
interested in the value of the spin $S \gg 1$, which translates into the
scale of orbital energies that is much larger than $\delta_1$. 
The fluctuations of the
density of states averaged over such an energy scale are much smaller
than its mean value $1/\delta_1$, and it is sufficient to
 keep only the terms linear in $\delta\rho$. Thus, we obtain
from Eq.~(\ref{IN4}):
\begin{eqnarray}
{\cal E}_{>}=S-\int_{0}^{S\delta_1}d\epsilon\delta\rho(\epsilon) ,
\nonumber\\
{\cal E}_{<}=-S+\int_{-S\delta_1}^{0}d\epsilon\delta\rho(\epsilon).
\label{AP}
\label{eq:3.3}
\end{eqnarray}
In what follows we adopt the notation:
\begin{equation}
\mu=2\delta_1 - 2J\ll \delta_1.
\label{eq:3.4}
\end{equation}
According to Eq.~(\ref{Stoner}), the bulk Stoner instability emerges at 
$\mu =0$,
so that the parameter $\mu$ characterizes how close to the criticality
threshold the system is.

Substituting Eqs.~(\ref{AP}) into Eq.~(\ref{IN3}), we obtain:
\begin{equation}
E(S,\xi)=\frac{\mu}{2}S^2-(\xi-1)(\delta_1-\mu/2)S-S\int^{S\delta_1}_{-S\delta_1} 
\delta\rho(\epsilon)d\epsilon
+\int^{S\delta_1}_0 (\delta\rho(\epsilon)+\delta\rho(-\epsilon))\epsilon 
d\epsilon,
\label{eq:3.5}
\end{equation}
Integrating Eq.~(\ref{eq:3.5}) by parts and using Eq.~(\ref{eq:3.4}) to
neglect $\mu$ as compared to $\delta_1$ we obtain
\begin{equation}
E(S,\xi)=\frac{\mu}{2}S^2-(\xi-1)\delta_1 S-\int^{S\delta_1}_0 
d\epsilon\int^{\epsilon}_0
d\epsilon_1\left[\delta\rho(\epsilon_1)+\delta\rho(-\epsilon_1)\right].
\label{es1}
\label{eq:3.6}
\end{equation}

Thus, we reduced the original problem to finding the minima
of the random function
\begin{equation}
E(S,\xi)=\frac{\mu}{2}S^{2}-(\xi-1)\delta_1 S+V(S),\label{M1}
\label{eq:3.7}
\end{equation}
where $\mu$ is fixed (and small compared to $\delta_1$), and $V(S)$ is a random potential
which is determined by  the fluctuations of the density of states
$\delta\rho(\epsilon)$
\begin{equation}
V(S)=-\int^{\delta_1 S }_0 d\epsilon\int^{\epsilon}_0
d\epsilon_1\left[\delta\rho(\epsilon_1)+\delta\rho(-\epsilon_1)\right].
\label{pot1}
\label{eq:3.8}
\end{equation}
Below we refer to this problem as a Random Potential Problem (RPP).
Such a problem is well defined provided that the correlation function
for random potential $V(S)$ is given.
One can evaluate the statistics of the potential $V(S)$,
Eq.~(\ref{eq:3.8}), using RMT.

The fluctuations of the
density of states, $\delta\rho$, averaged over energy intervals larger
than $\delta_1$ in the ensembles of RM are Gaussian random variables.
Therefore, using Eq.~(\ref{pot1}) we conclude that $V(S)$ is a Gaussian 
random variable as
well. The correlation function $\langle V(S)V(S')\rangle $ can be
expressed through the correlator of the density of states. The
correlation function of the Fourier components of
$\delta\rho(\epsilon)$ is given by\cite{10}:
\begin{equation}
\langle\delta\rho_k\delta\rho_p\rangle=\frac{2|k|}{\beta}\delta(p+k),
\label{AP3}
\end{equation}
where $\beta$ is $1$ or $2$ depending on whether the Hamiltonian belongs to
an orthogonal or a unitary ensemble. 
Averaging the product $V(S_1)V(S_2)$ with the help of 
Eqs.~(\ref{pot1}) and Eq.~(\ref{AP3}), we obtain the correlation function:
\begin{eqnarray}
\langle V(S_1)V(S_2)\rangle
=\frac{\delta_1^2}{\pi^2\beta}\left[-S_1^2\ln S_1^2-S_2^2\ln S_2^2+
\right.&&\label{M2}\\
+\left.\frac{(S_1+S_2)^2}{2}\ln (S_1+S_2)^2+\frac{(S_1-S_2)^2}{2}
\ln (S_1-S_2)^2 \right].&&\nonumber
\end{eqnarray}

Equations (\ref{eq:3.7}) and (\ref{M2}) define the random potential
problem. In the next section we will show that the statistical
description of this function possesses simple scaling properties.

\section{Scaling Analysis}
\label{sec:4}

It follows from Eq.~(\ref{M2}), that the correlation
function of the random potential is a homogeneous function of degree
two:
\begin{equation}
\langle V(\gamma S_1)V(\gamma S_2)\rangle =\gamma^2 \langle
V(S_1)V(S_2)\rangle.
\label{corr}
\label{eq:4.1}
\end{equation}
Equation (\ref{eq:4.1}) means that the probability of the potential
realization $V(S)$ is the same as the probability of the potential
$\gamma V(\gamma^{-1} S)$, and by no means expresses the scaling for the
potential of  a given realization.

Our goal now is to demonstrate, that
this property of the Gaussian random potential dictates
a particular scaling form for all the moments of the free energy
$F$ of the system:
\begin{eqnarray}
F\left(T,\mu; 
\left\{V(S)\right\}
\right) = -T \ln \left\{\sum_{S \geq 0 }
2\left(S+\frac{1}{2}\right)^{\xi -1}
\exp\left(-\frac{E(S,\xi)}{T}\right)\right\}\nonumber\\
= - T \ln \left\{2 
\int_0^\infty dS S^{\xi}
\exp\left(-\frac{E(S,\xi)}{T}\right)\right\},
\label{eq:4.2}
\end{eqnarray}
where energy $E(S)$ is given by Eq.~(\ref{eq:3.7}). Note
that in Eq.~(\ref{eq:4.2}), which is valid in the continuous limit, we
neglected unity in comparison with $2S$. This is because we have
already used several times the fact that $S \gg 1$, and keeping this
unity in the pre-exponential factor would be beyond the accuracy of the 
calculation.

It follows from Eqs.~(\ref{eq:4.2}) and (\ref{M1}) that for any given 
realization of the random
potential $V(S)$, the following identity holds:
\begin{equation}
F\left(\gamma T, \gamma^{-1}\mu;
\left\{\gamma V(\gamma^ {-1} S)\right\}
\right) + \gamma T\ln 2 + \gamma  \xi T\ln \gamma   = 
\gamma F\left(T, \mu; 
\left\{V(S)\right\}
\right) + T\ln 2.
\label{eq:4.3}
\end{equation}
According to Eq.~(\ref{eq:4.1}) the probability of the potential
realization $V(S)$ is the same as the probability of the potential
$\gamma V(\gamma^{-1} S)$. Therefore, the moments of the free energy 
possess the following scaling property
\begin{eqnarray}
\langle F\left(\gamma T, \gamma^{-1}\mu 
\right)\rangle + \gamma T\ln 2 + \gamma\xi T\ln \gamma   = 
\gamma \langle F\left(T, \mu
\right) 
\rangle
+ T\ln 2\nonumber\\
\langle F^n\left(\gamma T, \gamma^{-1}\mu 
\right)\rangle_c
= \gamma^n \langle F^n\left(T, \mu
\right)\rangle_c
,\quad n =2,3,4, \dots,
\label{eq:4.4}
\end{eqnarray}
where $\langle \dots \rangle$ stands for the ensemble averaging
and subscript ``$c$'' means the irreducible average (cumulant).

We can use the fact that the only available variable with the
dimensionality of energy is $\delta_1/\sqrt{\beta}$, 
see Eq.~(\ref{M2}), and conclude that 
the moments of the free energy should be of the following scaling form
\begin{eqnarray}
 \langle F\left(T, \mu 
\right) 
\rangle =
- T\left[\frac{\xi}{2}\ln \left(\frac{T}{\mu}\right) + \ln 2+
\frac{2-\xi}{2}\ln \frac{\pi}{2} \right]
- \frac{\delta_1^2}{\beta \mu}f^{(1)}_\xi\left(\theta, \beta\right);
\nonumber\\
\langle \left[F\left(T, \mu 
\right)\right]^n\rangle_c
=\left(
\frac{\delta_1^2}{\beta \mu}
\right)^{n}
f^{(n)}_\xi\left(\theta,\beta\right)
,\quad n = 2,3,\dots.
\label{eq:4.5}
\end{eqnarray}
Here we introduced scaling variable
\begin{equation}
\theta = \frac{\beta \mu T}{\delta_1^2},
\label{eq:4.6}
\end{equation}
that has the meaning of a dimensionless temperature.  In
Eq.~(\ref{eq:4.5}), $f^{(n)}_\xi$ are dimensionless functions
which can not be found from the scaling arguments alone.

The statistics of the magnetization can be calculated with the
help of the identity which follows straightforwardly from
Eqs.~(\ref{eq:3.7}) and (\ref{eq:4.2}):
\begin{equation}
\overline{S^{2}}= 2\frac{\partial F}{\partial \mu}.
\label{eq:4.7}
\end{equation}
The bar in Eq.~(\ref{eq:4.7}) and below stands for the 
thermodynamic average within a given realization of $V(S)$.
In a complete analogy with the derivation of Eq.~(\ref{eq:4.5}),
we obtain the following scaling behavior of the magnetization: 
\begin{eqnarray}
 \langle \overline{S^{2}}
\rangle =\frac{\xi T}{\mu} +
\frac{\delta_1^2}{\beta \mu^2}G^{(1)}_\xi\left(\theta, \beta\right);
\nonumber\\
\langle\left[
\overline{S^{2}}
\right]^n\rangle_c
=
\left(
\frac{\delta_1^2}{\beta \mu^2}
\right)^{n}
G^{(n)}_\xi\left(\theta, \beta\right),\quad n=2,3,\dots.
\label{eq:4.8}
\end{eqnarray}
For $n=1$ one finds from Eq.~(\ref{eq:4.7}) that
\begin{equation}
G^{(1)}_\xi(\theta, \beta) = -2 \theta^2 \frac{d}{d\theta}
\left[\theta^{-1}f^{(1)}_\xi (\theta, \beta)\right].
\label{eq:4.9}
\end{equation}
There is no  straightforward relation between $G^{(n)}$ and 
$f^{(n)}$ functions for $n>1$.
It is noteworthy that for the easy axis (Ising) case ($\xi=1$) the scaling 
functions $f^{(n)}_1\left(\theta,\beta\right)$ and 
$G^{(n)}_1\left(\theta,\beta\right)$ do not depend on $\beta$.

Let us discuss the asymptotic behavior of  functions ${\cal
F}^{(1,2)}_\xi$ from Eq.~(\ref{eq:4.5}). We begin with the high
temperature regime, $\theta \to \infty$.
One can expand Eq.~(\ref{eq:4.2}) up to the second order in the potential
$U(S)$, where
\begin{equation}
U(S)=-(\xi-1)S+V(S),
\end{equation}
to obtain
\begin{equation}
F\approx - T\frac{\xi}{2}\ln \left(\frac{T}{\mu}\right) - T\ln 2-
T\frac{2-\xi}{2}\ln \frac{\pi}{2}
+ \overline{\left. U(S) \right|_0} - 
\frac{ \overline{\left. U(S)^2 \right|_0} - 
\overline{\left. U(S) \right|_0}^2}{2 T}.
\label{eq:4.10}
\end{equation}
Here we introduced the following notation
\begin{equation}
\quad \overline{\left.\dots \right|_0}
\equiv \frac{\int_0^\infty dS S^{\xi-1} e^{-\mu S^2/2T} \dots}
{\int_0^\infty dS S^{\xi-1} e^{-\mu S^2/2T}}.
\label{eq:4.100}
\end{equation}
The fourth and fifth terms in the expansion (\ref{eq:4.10}) are random
 quantities. Averaging them  with the help of Eq.~(\ref{M2}), we obtain
\begin{mathletters}
\begin{equation}
f_1^{(1)}(\theta)=\frac{\ln 2}{\pi^2}+ {\cal O}(1/\theta);
\quad
f_1^{(2)}(\theta)=\theta \frac{2\ln 2}{\pi^2} + {\cal O}(1),
\label{eq:4.11}
\end{equation}
\begin{eqnarray}
f_2^{(1)}(\theta,\beta)
=\sqrt{\theta}\sqrt{\frac{\beta\pi}{2}}+\beta\left(2-\frac{\pi}{2}\right)+
\frac{4\ln 2}{\pi^2}-\frac{C_2}{2}
+ {\cal O}\left(\frac{1}{\sqrt{\theta}}\right);
\label{eq:4.11a}\\
f_2^{(2)}(\theta,\beta)
=\theta C_2+2\sqrt{\theta\beta}\left[D_2-\sqrt{\frac{\pi}{2}}C_2 
\right]+{\cal O}(1).\nonumber
\end{eqnarray}
\end{mathletters}
The numerical coefficients $C_2$ and $D_2$ have the following meaning
\begin{mathletters}
\begin{eqnarray}
C_2=\frac{\beta\mu}{T\delta_1^2}\langle \overline{\left. V(S) \right|_0}^2
\rangle;\nonumber\\
D_2=\beta\left(\frac{\mu}{T\delta_1^2}\right)^{\frac{3}{2}}\langle
\left(\overline{\left. V(S) \right|_0}\right)
\cdot \left(\overline{\left.S\cdot V(S) \right|_0}\right)
\rangle.\nonumber
\end{eqnarray}
\end{mathletters}
Their numerical values are 
\[
C_2=\frac{2}{\pi^2}\left(\int^1_0 \frac{dx}{\sqrt{x}}
\frac{(1-x)^2}{(1+x)^3}\ln \frac{1}{x}-1\right)=\frac{1}{4\pi^2}\left[
\Psi^{\prime}(1/4)-\Psi^{\prime}(3/4)\right]\approx 0.3712,
\]
\[
D_2=\pi^{-\frac{3}{2}}\left( 6\ln (\sqrt{2}+1)-2\sqrt{2}\ln 2 \right)
\approx 0.5976,
\]
where $\Psi^{\prime}(z)=\frac{d^2}{dz^2}\log \Gamma (z)$  is the second 
logarithmic derivative of the $\Gamma$-function.
Substituting Eqs.~(\ref{eq:4.11}) and (\ref{eq:4.11a})  into Eq.~(\ref{eq:4.9}), 
we find
\begin{equation}
G_1^{(1)}(\theta)=\frac{2\ln{2}}{\pi^2}+ {\cal O}
\left(\frac{1}{\theta}\right);
\quad
G_2^{(1)}(\theta)=2f_2^{(1)}(\theta,\beta)-\sqrt{\theta}
\sqrt{\frac{\beta\pi}{2}}+{\cal O}\left(\frac{1}{\sqrt{\theta}}\right).
\label{eq:4.110}
\end{equation}

Equations (\ref{eq:4.11}), (\ref{eq:4.11a}) and (\ref{eq:4.110}) are valid in the high temperature regime
where the disorder only weakly affects the temperature
fluctuations of the spin. There is no parametrically justified theoretical
approach to analyze the situation at low temperatures. The popular
approach to the RPP is the Replica Symmetry
Breaking ansatz\cite{3,Mezard91,Mezard92,Engel}. 
The results of such a calculation will be published elsewhere\cite{elsewhere}. 
Here we restrict ourselves
to a qualitative consideration, which yields the answers up to
numerical coefficients.

We employ arguments similar
to those of Larkin\cite{Larkin} for the collective pinning of the vortex 
lattice, and of
Imry and Ma\cite{7}  for the random spin systems. Let us first discuss the 
Ising, $\xi=1$, case.
At the point $S_g$ of the global minimum  of the energy $E(S)$,
Eq.~(\ref{M1}),
${\mathrm min}\{E(S)\}=E(S_g)$ the
random potential $V(S_g)$ is of the same order of magnitude as the quadratic
term $\mu S_g^2/2$ and has the opposite sign. This condition can be written
as
\begin{equation}
\frac{\mu}{2}S^{2}_g\simeq \sqrt{\langle V^2(S_g)\rangle}.
\label{eq:4.12}
\end{equation}
At $T=0$ the entropy term is not important. Hence the spin at zero
temperature 
is equal to $S_g$ and  does not depend on the degeneracy of the state.
Using Eq.~(\ref{M2}), we find that
\begin{equation}
\langle \overline{S^2}\rangle^I \simeq \frac{\delta_1^2}{\beta\mu^2}.
\label{eq:4.13}
\label{lowT}
\end{equation}
Therefore, the function $G_1^{(1)}$ tends to a constant independent of 
$\beta$ as $\theta 
\to 0$. 

The estimate of the position of the minimum, $S_g$, 
for the Heisenberg case, $\xi=2$, differs from Eq.~(\ref{eq:4.12}) only 
slightly. It follows from Eq.~(\ref{M1}) that
\begin{equation}
\mu S_g-1 \approx \sqrt{\langle V^2(S_g)\rangle} .\label{nodisorder}
\end{equation}
As a result
\begin{equation}
\langle \overline{S^2}\rangle^H= \frac{\delta_1^2}{\mu^2}
\left(1+\frac{a}{\beta}\right).\label{s2h}
\end{equation}
Here $a$ is a numerical constant of order unity, which cannot be determined 
within scaling considerations only. Therefore, the function 
$G_2^{(1)}$ tends to a constant as $\theta \to 0$ but in the 
Heisenberg case this constant depends on $\beta$.

According to Eq.~(\ref{eq:4.9}), this means that
$f^{(1)}$ also has a finite limit at $T=0$.
Moreover, the squared spin of the ground state fluctuates
from sample to sample and the  fluctuations are of the order of its
average value ${\langle S_g^2\rangle}$. 
As a result,  all of the functions $f^{(n)}, \ {\cal
G}^{(n)} $ must reach finite limits for both Heisenberg and Ising cases.

Let us discuss  the physical meaning of the
low-temperature expansion using 
\begin{equation}
G_\xi^{(1)}(\theta) = G_\xi^{(1)}(0 ) + 
\theta\frac{d G_\xi^{(1)}}{d\theta} + {\cal O}\left(\theta^2\right).
\label{eq:4.14}
\end{equation}
as an example.
Both the constant term and the derivative have a simple interpretation in
terms of the characteristics of the absolute minima of the random
potential $V(S)$.

Consider a realization of the random potential $V(S)$. The energy
of the system (\ref{eq:3.7}) has only a finite number of minima and
there are no symmetry reasons for degeneracies. Therefore, at low enough
temperatures only the vicinity of the ground state spin $S_g$
determines all of the thermodynamic properties of the system.
Close to the minimum the Hamiltonian can be approximated as
\begin{equation}
E(S) \approx E(S_g) + \frac{\mu_r}{2}\left(S-S_g\right)^2.
\label{eq:4.15}
\end{equation}
The corresponding value of the square of the magnetization is
\begin{equation}
\overline{S^2}= S_g^2 + 3^{\xi-1}\frac{T}{\mu_r}.
\label{eq:4.16}
\end{equation}
Naturally, $\overline{S^2}$ is a sample dependent quantity as well as
$\mu_r$ and $S_g$. Averaging (\ref{eq:4.16}) over the realization
and comparing the result with Eqs.~(\ref{eq:4.8}), (\ref{eq:4.14}),
we find
\begin{eqnarray}
\langle {S_g}^2
\rangle
=\frac{\delta_1^2}{\beta \mu^2}
G^{(1)}_\xi\left(0\right),
\nonumber\\
3^{\xi -1 }
\mu\left\langle\frac{1}{\mu_r}\right\rangle
= \frac{d G_\xi^{(1)}}{d\theta} + \xi.
\label{eq:4.17}
\end{eqnarray}
This means that the low-temperature expansion determines the averaged
location of the absolute minima and the curvature in this
minima. 
\section{Numerical Simulations}
\label{sec:5}

The RPP (\ref{eq:3.7})
with the Gaussian random potential $V(S)$, characterized by its correlator
(\ref{M2}), is easily accessible for numerical simulations.  Indeed,
instead of generating the ensemble of potentials V(S) with the
given correlator one can use the connection of V(S) with the spectra
of random matrices.  [See Sec.~\ref{sec:3}]. 

We carried out numerical simulations for both the orthogonal ($\beta=1$)
and unitary ($\beta=2$) ensembles.  In the former case we generated
$1200 \times 1200$ symmetric ($H_{\alpha\gamma} = H_{\gamma\alpha}$)
matrices with real matrix elements $H_{\alpha\gamma}$.  These matrix
elements were independent Gaussian random numbers with the following
 moments:
\begin{equation}
\langle H_{\alpha\gamma} \rangle=0;
\quad
\langle (H_{\alpha\gamma})^2 \rangle=1 \label{rmdisp}.
\end{equation}

To obtain a
matrix from the unitary ensemble we generated simultaneously a
symmetric ${\mathrm Re} H_{\alpha\gamma}$ and an antisymmetric ${\mathrm
Im} H_{\alpha\gamma}$ real matrices (${\mathrm Re}H_{\alpha\gamma} =
{\mathrm Re} H_{\gamma\alpha}$; ${\mathrm Im} H_{\alpha\gamma} = -
{\mathrm Im} H_{\gamma\alpha}$) with the same dispersion as above 
(Eq.~(\ref{rmdisp})).  The combination ${\mathrm
Re}H_{\alpha\gamma}+i {\mathrm Im} H_{\alpha\gamma}$ is a matrix
element of the Hamiltonian from the unitary ensemble.

In our analytic calculations we assumed that the mean level spacing
$\delta_{1}$ does not depend on the location of the energy strip where
$\delta_{1}$ is calculated.  Strictly speaking,  this is not the case
for the Gaussian ensembles of random matrices.  It is well known
\onlinecite{11}, that the density of the Random Matrix
eigenvalues is distributed according to the Wigner semi-circle law: this
density is larger in the middle of the band ($\epsilon$ close to zero)
and vanishes at the band edges $\pm \epsilon_{0}$ as
$\sqrt{\epsilon^2_{0} - \epsilon^{2}}$.  Accordingly the mean level
spacing depends on the energy:
\begin{equation}
\delta_{1} (\epsilon) = \delta_{1}(0) 
{\sqrt{\frac {\epsilon^{2}_{0}}{\epsilon^{2}_{0} -
      \epsilon^{2}}}}.
\label{eq:5.1}
\end{equation}
It is also well known that Eq.~(\ref{eq:5.1}) is just
an asymptotic law, which becomes exact in the limit $N
\rightarrow \infty$, where N is the rank of the matrices;
$\epsilon_{0} \propto {\sqrt N}$.  At finite $N$ there are corrections to
Eq.~(\ref{eq:5.1}), which become most pronounced close to the edges\cite{11}.

Taking all of this into account we first discarded the lowest and the
highest $100$ states in the spectrum of each Random Matrix.  After
that, we unfolded the rest of the spectrum according to
Eq.~(\ref{eq:5.1}) and obtained for each matrix $1000$ eigenstates that
obey local Wigner-Dyson statistics and have uniform density.  We also
scaled out $\delta_1(0)$ and ended up with the mean level spacing equal
to unity.

To evaluate the realization of $V(S)$ which corresponds to a given
Random Matrix one can simply sum up energies of the lowest $500+S$
states (filled by electrons with spin up) and the lowest $500-S$
states (spins down). Subtracting $S^{2}$ from the resulting sum we
obtain the Random Potential $V(S)$ in units of $\delta_1$.  
Some particular realization of the random potential for different
$\beta$ are presented on Fig.~\ref{fig:5.0}.

Using the generated potential $V(S)$, we calculated the free
energy, $F(\mu,T)$, Eq.~(\ref{eq:4.2}) and the thermodynamic magnetization
\begin{eqnarray}
\displaystyle{
\overline{S^2}= \frac{1}{Z}\sum_{S \geq 0 }
S^2 \left(2 S+{1}\right)^{\xi -1} 
\exp\left(-\frac{E(S)}{T}\right);}
\label{eq:5.200}\\ 
\displaystyle{ 
Z=\sum_{S \geq 0 }
\left(2 S+{1}\right)^{\xi -1} 
\exp\left(-\frac{E(S)}{T}\right)
}
\end{eqnarray}
for a given realization
and then evaluated the ensemble average of different moments of random 
$F(\mu,T)$ and $\overline{S^2}$.
All of the data presented below are results of averaging over $12000$ 
realizations of random matrices.

Figure~\ref{fig:5.1} demonstrates the scaling properties
for the mean free energy  derived in the
preceding section, Eq.~(\ref{eq:4.5}).
We evaluated $f^{(1)}$
for different values of $T,\ \mu$ and $\beta$, and plotted it as a
function of the scaling variable $\theta$, Eq.~(\ref{eq:4.6}).
One can see that for the Ising case the
 data for different values of $\mu$ and for both $\beta=1$ and $\beta=2$
collapses on a single curve in accord with Eq.~(\ref{eq:4.5}).
The scaling function $f_1^{(1)}(\theta)$ at $\theta\gg 1$ approaches
its high-temperature asymptotic 
value of $\pi^{-2}\ln 2\approx 0.070$ (Eq.~(\ref{eq:4.11})) within statistical
errors. 

For the Heisenberg case the high temperature expansion predicts that 
$f_2^{(1)}$ behaves at $\theta\gg 1$ as 
(compare with Eq.~(\ref{eq:4.11a}))
\begin{equation}
f_2^{(1)}(\theta,\beta)
\approx 1.25\sqrt{\theta\beta}+0.429\beta+
0.095.\label{f1ass}
\end{equation}
The best fit lines obtained for the numerical data are described by
\begin{equation}
f_2^{(1)}(\theta,\beta)
\approx (1.182\pm 0.002)\sqrt{\theta\beta}+(0.31\pm 0.02)\beta+
0.12\pm 0.03.
\end{equation}
This is in a good agreement with Eq.~(\ref{f1ass}) taking into 
account the fact that there should be corrections to Eq.~(\ref{eq:4.11a}) of
 order $1/\sqrt{\theta}$ and
$\mu$ (since we neglected $\mu$ as compared to $\delta_1$ using the 
condition in Eq.~(\ref{eq:3.4})).

Figure~\ref{fig:5.2} illustrates the behavior of the mesoscopic fluctuations
of the free energy, see Eq.~(\ref{eq:4.4}),
\[
f_\xi^{(2)}(\theta)
= \frac{(\beta\mu)^2}{\delta_1^4}
\left(\langle\left[{F}(T,\mu)\right]^2\rangle -
\langle{F}(T,\mu)\rangle^2\right)
.
\]

For the Ising case at $\theta \gg 1$ according to Eq.~(\ref{eq:4.11}) 
$f_1^{(2)}\approx 0.1405\cdot \theta$ . The numerical simulations give 
the slope of the best 
fit line equal to $0.1358\pm 0.0002$. Therefore, the results agree well.
 We can rewrite Eq.~(\ref{eq:4.11a}) for the Heisenberg case as  
\begin{equation}
\frac{f_2^{(2)}(\theta,\beta)}{\sqrt{\theta}}
\approx 0.371\sqrt{\theta}+2\cdot 0.132\sqrt{\beta}+{\cal O}(1/\sqrt{\theta}),
\end{equation}
i.e $f_2^{(2)}(\theta,\beta)/\sqrt{\theta}$ is a linear function of 
$\sqrt{\theta}$. The numerics indeed demonstrates such a linear dependence 
which can be best fitted by
\begin{equation}
\frac{f_2^{(2)}(\theta,\beta)}{\sqrt{\theta}}
\approx (0.353\pm0.006)\sqrt{\theta}+2\cdot (0.160\pm 0.003)\sqrt{\beta}+{\cal O}(1/\sqrt{\theta}).
\end{equation}
Once again, the agreement is quite reasonable since there are corrections 
${\cal O}(\mu)$ to the coefficients.

Figure~\ref{fig:5.3} presents the numerical results for the spontaneous
magnetization (\ref{eq:5.200}). We plot the
difference between $\langle \overline{S^2} \rangle$ and its
high-temperature asymptotic:
\begin{equation}
G^{(1)} = \mu^2\beta \left[\langle \overline{S^2} \rangle
-  \frac{\xi T}{\mu}\right]
\end{equation}
as the function of the scaling variable $\theta$. Once again, all the
curves collapse in accordance with Eq.~(\ref{eq:4.8}). The function 
$G^{(1)}_1$ 
approaches the asymptotic value within statistical errors. 
The collapse of the 
data at $\theta \rightarrow 0$ justifies the order of magnitude estimate 
that led to Eq.~(\ref{eq:4.13}).
For the Heisenberg case 
Eq.~(\ref{eq:4.110}) predicts the following high $\theta$ behavior
\begin{equation}
G_2^{(1)}(\theta)=1.25\sqrt{\theta\beta}+0.858\beta+0.190
+{\cal O}\left(\frac{1}{\sqrt{\theta}}\right).
\end{equation}
The best fit lines are described by
\begin{equation}
G_2^{(1)}(\theta)=(1.171\pm 0.006)\sqrt{\theta\beta}+
(0.54\pm 0.01)\beta+(0.29\pm 0.03).
\end{equation}
Again the results agree up to ${\cal O}(\mu)$ in the slope and 
${\cal O}\left(1/\sqrt{\theta}\right)$ in the intercept. 
There are downward deviations for the smallest $\mu$ at high 
temperature. They are likely due to the finite size effects (the magnetization 
becomes too close to its maximal value $S=500$ determined by the size of the 
RM). As a result the scaling is violated. 

As a matter of fact, the scaling is violated when $\mu$ is too small or
too large. At 
large $\mu$ the typical ground state spin becomes of the 
order of unity and the 
condition $S\gg 1$, used throughout this article, no longer holds. As $\mu$
decreases the magnetization becomes of the order 
of the system size, 
$S\approx 500$, whereas in making arguments about scaling we assumed no upper 
bound on the value of $S$. Therefore, in obtaining the numerical values for 
the scaling functions at $\theta \to 0$ we used the values of $\mu$ that 
correspond to the ground state magnetization $S_g$ from $\sim 15$ to $\sim 150$
hundreds. The linear interpolation of the numerical curves at $\theta\to 0$
results in
the following values for the low temperature asymptotics of the scaling 
functions $G^{(1)}$, Eq.~(\ref{eq:4.8})
\begin{equation}
G^{(1)}_{1}\left(0\right)
=.256\pm .005;\quad G^{(1)}_{2}(0,\beta=1)=1.60\pm .01;
\quad G^{(1)}_{2}(0,\beta=2)=2.65\pm .01.
\label{eq:5.4a}
\end{equation}
In the Ising case the slope of this function can be determined rather well
\begin{mathletters}
\begin{equation}
{ \frac{d G_{1}^{(1)}}{d\theta} }=-0.7\pm 0.1,\label{eq:5.4i}
\end{equation}
whereas evaluation of this slope in the Heisenberg case requires much better
statistics. From what we had it follows that
\begin{equation}
{ \frac{d G_{2}^{(1)}(\beta=1)}{d\theta} }=0.1\pm 0.2;
\quad { \frac{d G_{2}^{(1)}(\beta=2)}{d\theta} }=0.2\pm 0.3.
\label{eq:5.4b}
\end{equation}
Even though the values of the slope are smaller than the statistical errors
we do know the behavior of $S^2$ at low temperature. The smallness of the slope
just means that the change of magnetization squared 
with temperature is only slightly different from the one 
predicted by the high $T$ expansion.   
\end{mathletters}
Using Eqs.~(\ref{eq:5.4i}), (\ref{eq:5.4b})  and Eqs.~(\ref{eq:4.17}), 
we conclude that
\begin{mathletters}
\begin{eqnarray}
\langle {S_g}^2
\rangle ^I
=(.256\pm .005) \frac{\delta_1^2}{\beta \mu^2},\\
\langle {S_g}^2
\rangle^H_{\beta=1}
=(1.60\pm 0.01) \frac{\delta_1^2}{\mu^2};\quad\langle {S_g}^2
\rangle^H_{\beta=2}=(1.33\pm 0.01)\frac{\delta_1^2}{\mu^2},
\end{eqnarray}
\end{mathletters}
and
\begin{equation}
\left\langle\frac{1}{\mu_r}\right\rangle
=\frac{1}{\mu} \left\{
\matrix{0.3\pm0.1, & \xi=1\cr
0.7\pm0.1, & \xi=2\cr}
\right.
\label{eq:5.5}
\end{equation}

Figure \ref{fig:5.4} illustrates the behavior of the averaged 
and rescaled zero temperature 
magnetization $\mu\langle\left|S_g\right|\rangle^H$ and its 
square $\mu^2\langle S_g^2 \rangle^H$ in the Heisenberg case. The 
magnetization squared $\langle S_g^2 \rangle^H$ is well described by 
Eq.~(\ref{s2h}) with the numerical constant $a\approx 0.6$. 
The contribution from disorder to the averaged 
magnetization $\langle\left|S_g\right|\rangle^H$ is an order of magnitude 
smaller
than the magnetization itself. Without disorder ($V(S)=0$) the magnetization 
is the same for all grains $S_g=1/\mu$ (Eq.~(\ref{M1})). 
The correction due to 
randomness is around $9\%$ in the orthogonal $\beta=1$ and $4\%$ in the 
unitary $\beta=2$ case.
 
\section{Conclusions}


We considered manifestations of electron-electron interactions in the 
properties 
of isolated metallic grains with large Thouless conductance, $g\gg 1$. It 
turned out that the interaction effects can be taken into account by a rather 
simple interaction Hamiltonian~(\ref{eq:2.17}). We then applied this 
description to study the mesoscopic spontaneous magnetization of the metallic 
grains whose bulk counterparts are only slightly below the point of Stoner 
instability. In this case the problem maps onto the random potential problem,
Eq.~(\ref{M1}), where the total spin of the system plays the role of the coordinate.
The randomness is manifested by the potential $V(S)$ and is entirely due
to the fact that {\em the one-electron spectrum} in such a grain
is sample specific. At the same time, the fluctuations of the exchange
interaction constant can be neglected provided that $g \gg 1$. 

The correlation function (\ref{M2}) of the random potential $V(S)$
follows directly from the Wigner-Dyson spectral statistics and
possesses a specific invariance (\ref{eq:4.1}) under scaling
transformations.
This invariance dictates a particular scaling of the ensemble
averaged thermodynamic properties of the grains as well as of the
higher moments of their mesoscopic fluctuations. Dependence of all
these quantities on temperature $T$ and on the distance from the point
of Stoner instability $\mu$ can be determined, see Eqs.~(\ref{eq:4.5}),
(\ref{eq:4.8}), and (\ref{eq:4.9}), up to some functions of the
dimensionless effective temperature $\theta = \beta\mu T$, where
$\beta=1(2)$ corresponds to the orthogonal (unitary) Dyson ensemble.

According to Eq.~(\ref{eq:4.8}), in the Ising case the zero-temperature 
magnetization
typically gets reduced by a factor of $\sqrt{2}$ when the system is driven
from $\beta=1$ to $\beta=2$. In the Heisenberg case the average 
zero-temperature magnetization is largely determined by the non-random part 
of the 
Hamiltonian~(\ref{M1}) (without disorder $S_g=1/\mu$). 
The fluctuations of the magnetization become
suppressed by a factor of $\sqrt{2}$ as the system goes from the 
orthogonal to the unitary ensemble. If the grains are large enough, the
transition between these ensembles can be completed in
magnetic fields, which produce still negligible Zeeman splitting. As a
result, an anomalously weak magnetic field would substantially
{\em reduce} the spontaneous magnetization in the Ising case or suppress its 
fluctuations in the Heisenberg one. This is due to the
well-known fact that the unitary spectra are more rigid than
orthogonal ones. However, the difference 
between the average magnetization 
with and without disorder is about 9\% for orthogonal ($\beta=1$) and 4\% for 
unitary ($\beta=2$) ensembles (see Fig.~\ref{fig:5.4}) in the Heisenberg 
case. Therefore, in 
the Heisenberg case a small magnetic field should suppress the average 
magnetization by only 5\%. This is a much smaller effect than the one 
predicted in 
Ref. \cite{14}. The discrepancy is due to erroneous choice of the model 
Hamiltonian
in that reference (see the discussion in Section II after 
Eq.~(\ref{doublediagonal})). 

Of course, the evaluation of the scaling
functions lies beyond the simple analysis. In the
high-temperature regime it is possible to develop a regular
perturbative expansion. At low temperature an analytic technique based
on Replica Symmetry breaking paradigm can be used. The corresponding
calculation will be reported elsewhere\cite{elsewhere}.  

In the
present paper, we analyzed the low-temperature asymptotic behavior
 numerically.
We have shown that these asymptotics are determined by a single absolute
minimum of a random potential (not accessible by a regular
perturbation theory). Using those numerical results we were able to
determine the average position and curvature for such minima, see 
Eqs.~(\ref{eq:5.4a}) and (\ref{eq:5.4b}).

\acknowledgements
We are grateful to Yu. M. Galperin, L.B. Ioffe, M. Mezard,  G. Parizi,
and R. Berkovits 
for helpful discussions.  
I.A. was supported by Packard research foundation fellowship.
The work at Princeton University was supported by ARO MURI DAAG
55-98-1-0270 and by NSF MRSEC grant DMR 98-09483.

{\begin{figure}[ht]
\epsfxsize=8cm
\centerline{\epsfbox{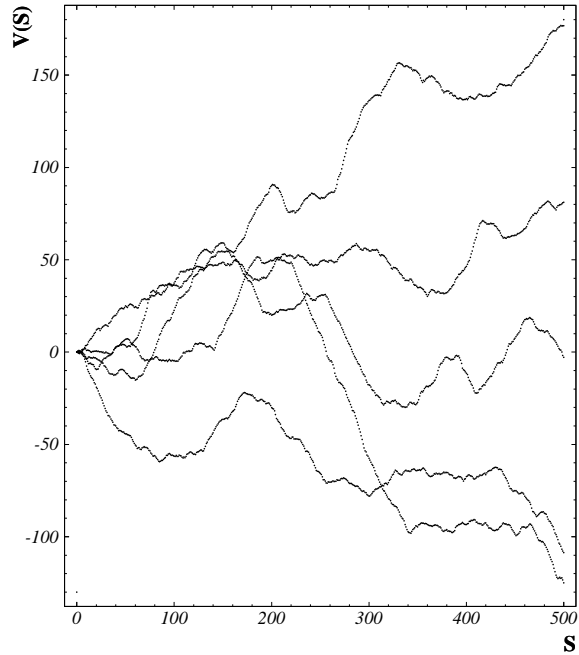}}
\vspace{-1cm}
\caption{Several realizations of the random potential $V(S)$ for $\beta=2$.
}
\label{fig:5.0}
\end{figure} 
}

{\begin{figure}[ht]
\epsfxsize=10cm
\centerline{\epsfbox{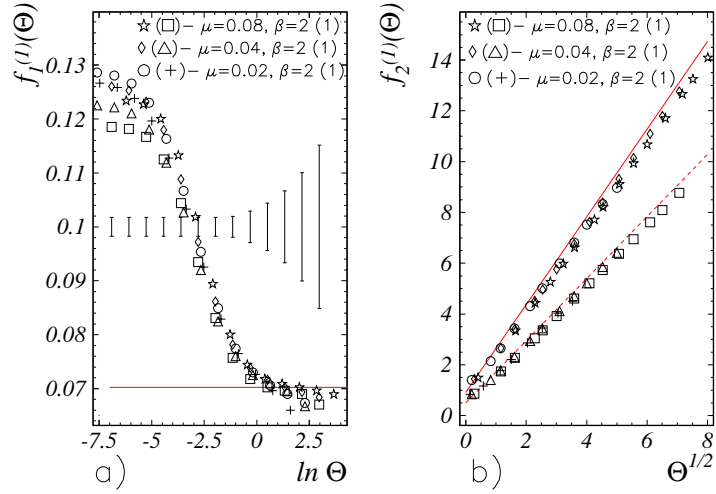}}
\caption{
a) The contribution to the free energy from disorder (function 
$f_1^{(1)}$, Eq.~(\ref{eq:4.5})) in the Ising ($\xi=1$) 
case. The logarithmic 
scale is chosen to demonstrate the fact that the scaling function
goes to a constant in both low and high temperature regimes. The solid line is
the result of the high temperature expansion. The statistical errors depend 
on $\theta$ only and are plotted in the center of the Figure.
b) Function $f_2^{(1)}$ in the Heisenberg 
case ($\xi=2$). The scaling functions are different for the unitary 
($\beta=2$) and 
orthogonal ($\beta=1$)ensembles. The $\sqrt{\theta}$ scale was chosen to 
illustrate the agreement with the high
temperature expansion obtained in Eq.~(\ref{eq:4.11a}). Solid (dotted) line 
represents the predicted high $\theta$ behavior for the unitary (orthogonal) 
case. At low 
temperature the functions tend to constants larger than the ones predicted by 
the high $\theta$ expansion.
}\label{fig:5.1}
\end{figure} 
}

{\begin{figure}[ht]
\epsfxsize=10cm
\centerline{\epsfbox{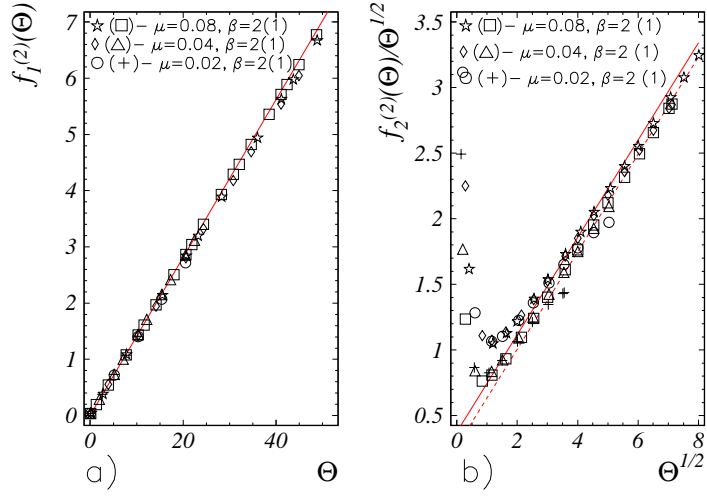}}
\caption{ a) The averaged fluctuations of the free energy of the 
grain  in the Ising case ($\xi=1$) rescaled according to 
Eq.~(\protect\ref{eq:4.5}). At $\theta\rightarrow 0$ the scaling function 
approaches a positive constant. The solid line is the high $\theta$ asymptotic
behavior 
described by Eq.~(\ref{eq:4.11}). 
b) The averaged fluctuations of the free 
energy in the Heisenberg case rescaled as in a). We divided the scaling 
function $f_2^{(2)}$ by $\sqrt{\theta}$ and plotted the ratio as 
a function of $\sqrt{\theta}$ to 
demonstrate the agreement with the high temperature expansion 
(Eqs.~(\protect\ref{eq:4.11a})). The solid (dotted) line 
represents high $\theta$ asymptotic behavior for the unitary (orthogonal) case 
$\beta=2(1)$. 
}\label{fig:5.2}
\end{figure} 
}

{\begin{figure}[ht]
\vspace{0.2cm}
\epsfxsize=10cm
\hspace*{0.5cm}
\centerline{\epsfbox{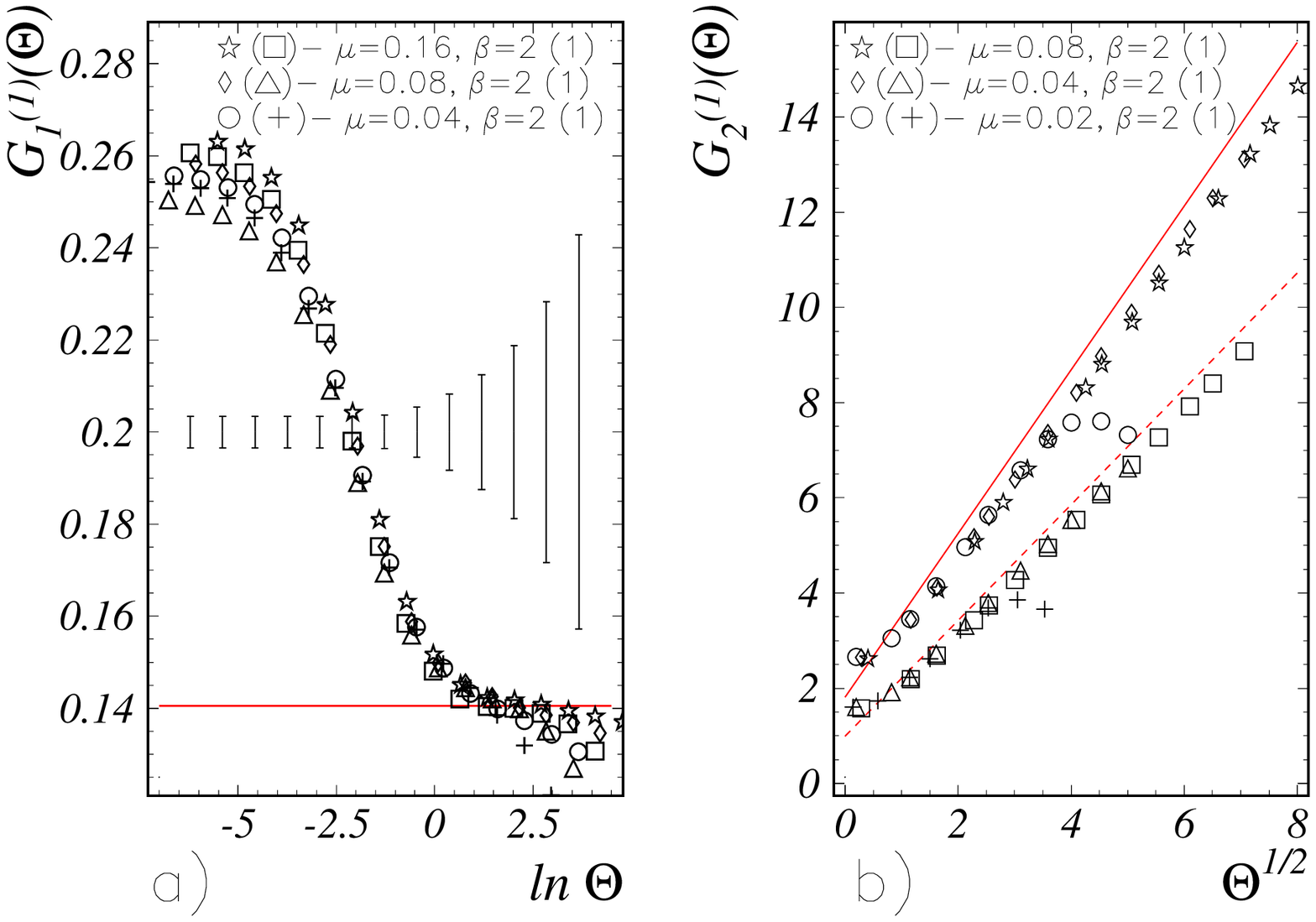}}
\caption{ a) The contribution from disorder to the averaged magnetization squared
in Ising case ($\xi=1$) scaled by $\beta\mu^2$
(see Eq.~(\protect\ref{eq:4.8}) for the definition). One can see that the 
scaling function $G_1^{(1)}$ has constant limits in both low and high 
temperature regimes. The solid line is
a high temperature expansion (Eq.~(\ref{eq:4.110})).The statistical errors 
depend 
on $\theta$ only and are plotted in the center of the Figure. 
b) The contribution from disorder and $-JS$ term to the averaged 
magnetization squared in the Heisenberg 
case ($\xi=2$) scaled as in a). The scaling functions are different for the 
unitary ($\beta=2$) and 
orthogonal ($\beta=1$)ensembles. The $\sqrt{\theta}$ scale was chosen to 
illustrate the agreement with the high
temperature expansion obtained in Eq.~(\ref{eq:4.110}). Solid (dotted) line 
represents result of high $\theta$ expansions for the unitary (orthogonal) 
case. At low 
temperature the functions go to constants higher than the ones predicted by 
the high $\theta$ expansions.}\label{fig:5.3}
\end{figure}}
{\begin{figure}[ht]
\vspace{0.2cm}
\epsfxsize=8cm
\hspace*{0.5cm}
\centerline{\epsfbox{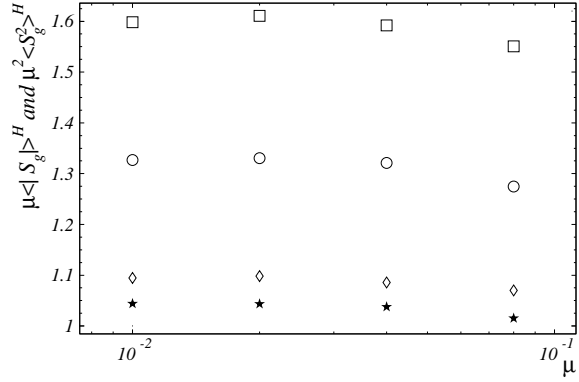}}
\caption{$\Box$ ($\bigcirc$) represents $\langle \mu^2 S_g^2 \rangle^H$ for 
$\beta=1(2)$. 
$\Diamond$ ($\star$) represents $\langle \mu |S_g| \rangle^H$ 
for $\beta=1(2)$. 
All data is for the Heisenberg ($\xi=2$) case.
Without disorder the relation  $\langle \mu^2 S_g^2\rangle=\langle 
\mu |S_g| \rangle=1$ holds. ($\delta_1$ is set to $1$).
}
\label{fig:5.4}
\end{figure} 
}
\end{document}